\documentclass{emulateapj}
\usepackage{graphicx}
\usepackage[usenames,dvips]{color}
\def\arcsec{$\,^{\prime\prime}$~}

\def\deg{$^{\circ}$~}

\newcommand{\be}{\begin{equation}}
\newcommand{\bel}[1]{\begin{equation}\label{eq:#1}}
\newcommand{\ee}{\end{equation}}
\newcommand{\bd}{\begin{displaymath}} 
\newcommand{\ed}{\end{displaymath}}   
\newcommand{\bea}{\begin{eqnarray}}
\newcommand{\beal}[1]{\begin{eqnarray}\label{eq:#1}}
\newcommand{\eea}{\end{eqnarray}}

\newcommand{\eqref}[1]{\ref{eq:#1}}

\newcommand{\lsim }{{\lower0.8ex\hbox{$\buildrel <\over\sim$}}}
\newcommand{\gsim }{{\lower0.8ex\hbox{$\buildrel >\over\sim$}}}

\def\Chandra{${\it Chandra}$}
\def\RXTE{${\it RXTE}$}
\def\Swift{${\it Swift}$}
\def\HST{${\it HST}$}
\def\ergss{ergs s$^{-1}$}

\def\simge{\mathrel{%
   \rlap{\raise 0.511ex \hbox{$>$}}{\lower 0.511ex \hbox{$\sim$}}}}
\def\simle{\mathrel{
   \rlap{\raise 0.511ex \hbox{$<$}}{\lower 0.511ex \hbox{$\sim$}}}}

\newcommand{\Msun}{\ifmmode {M_{\odot}}\else${M_{\odot}}$\fi}
\newcommand{\Lsun}{\ifmmode {L_{\odot}}\else${L_{\odot}}$\fi}
\newcommand{\Rsun}{\ifmmode {R_{\odot}}\else${R_{\odot}}$\fi}

\shorttitle{New transient LMXB in NGC 6440}
\shortauthors{Heinke et al.}

\begin{document}
\title{Discovery of a Second Transient Low-Mass X-ray Binary in the Globular Cluster NGC 6440}  

\author{C.~O. Heinke\altaffilmark{1}, D. Altamirano\altaffilmark{2}, H.~N. Cohn\altaffilmark{3}, P.~M. Lugger\altaffilmark{3}, S.~A. Budac\altaffilmark{1}, M. Servillat\altaffilmark{4}, M. Linares\altaffilmark{5,2}, T.~E. Strohmayer\altaffilmark{6}, C.~B. Markwardt\altaffilmark{6}, R. Wijnands\altaffilmark{2},  J.~H. Swank\altaffilmark{6}, C. Knigge\altaffilmark{7}, C. Bailyn\altaffilmark{8}, J.~E. Grindlay\altaffilmark{4} }

\altaffiltext{1}{Dept. of Physics, University of Alberta, Room 238 CEB, Edmonton, AB T6G 2G7, Canada; heinke@ualberta.ca}
\altaffiltext{2}{Astronomical Institute 'Anton Pannekoek', University of Amsterdam, Science Park 904, 1098 XH, Amsterdam, Netherlands}
\altaffiltext{3}{Dept. of Astronomy, Indiana University, 727 East Third St., Bloomington IN 47405}
\altaffiltext{4}{Harvard College Observatory, 60 Garden Street, Cambridge MA 02138}
\altaffiltext{5}{MIT Kavli Institute for Astrophysics \& Space Research, 70 Vassar St., Cambridge MA 02139}
\altaffiltext{6}{Astrophysics Science Division, NASA/GSFC, Greenbelt, MD 20771}
\altaffiltext{7}{School of Physics and Astronomy, University of Southampton, Hampshire SO17 1BJ, UK}
\altaffiltext{8}{Dept. of Astronomy, Yale University, PO Box 208101, New Haven, CT 06520}


\begin{abstract}
We have discovered a new transient low-mass X-ray binary, NGC 6440 X-2, with \Chandra/ACIS, \RXTE/PCA, and \Swift/XRT observations of the globular cluster NGC 6440.  The discovery outburst (July 28-31, 2009) peaked at $L_X\sim1.5\times10^{36}$ \ergss, and lasted for $<$4 days above $L_X=10^{35}$ \ergss.  
Four other outbursts (May 29-June 4, Aug. 29-Sept. 1, Oct. 1-3,  and Oct. 28-31 2009) have been observed with \RXTE/PCA (identifying millisecond pulsations, Altamirano et al. 2009a) and \Swift/XRT (confirming a positional association with NGC 6440 X-2), with similar peak luminosities and decay times. 
Optical and infrared imaging did not detect a clear counterpart, with best limits of $V>21$, $B>22$ in quiescence from archival \HST\ imaging, $g'>22$ during the August outburst from Gemini-South GMOS imaging,  and $J\gsim18.5$ and $K\gsim17$ during the July outburst from CTIO 4-m ISPI imaging.  
 Archival \Chandra\ X-ray images of the core do not detect the quiescent counterpart ($L_X<1-2\times10^{31}$ \ergss), and place a bolometric luminosity limit of $L_{NS}< 6\times10^{31}$ \ergss\ (one of the lowest measured) for a hydrogen atmosphere neutron star.  A short \Chandra\ observation 10 days into quiescence found two photons at NGC 6440 X-2's position, suggesting enhanced quiescent emission at $L_X\sim6\times10^{31}$ \ergss.
  NGC 6440 X-2 currently shows the shortest recurrence time ($\sim$31 days) of any known X-ray transient, although regular outbursts were not visible in the bulge scans before early 2009.  Fast, low-luminosity transients like NGC 6440 X-2 may be easily missed by current X-ray monitoring.

\end{abstract}


\maketitle

\section{Introduction}\label{s:intro}

The dense cores of globular clusters are known to be efficient factories for dynamically producing tight binaries containing heavy stars, and thus X-ray binaries \citep{Clark75,Hut91,Pooley03}.  Thirteen luminous ($L_X>10^{35}$ \ergss) low-mass X-ray binaries (LMXBs) have previously been identified in Galactic globular clusters \citep[see][]{Verbunt04}, concentrated in the densest and most massive clusters \citep{Verbunt87,Verbunt02}.  Luminous LMXBs in globular clusters of other galaxies are clearly concentrated in the most massive, most metal-rich, and densest globular clusters \citep{Kundu02,Sarazin03,Jordan04}.  A critical question for studies of globular cluster LMXBs is whether X-ray emission from a globular cluster is due to one LMXB or multiple LMXBs.  This affects the inferred nature of sources \citep{Dotani90,diStefano02,Maccarone07} and luminosity functions \citep{Sivakoff07}.  For example, an apparent contradiction in the qualities of the LMXB in M15 was resolved by the identification of two persistent X-ray sources in the cluster \citep{White01}.  Identification with \Chandra\ of multiple quiescent LMXBs in several globular clusters \citep{Grindlay01a,Pooley02b,Heinke03d} has suggested that transient LMXB outbursts from a cluster might arise from different sources, although some LMXBs possess distinctive characteristics \citep[e.g. the Rapid Burster,][]{Homer01}.
Here we report the discovery and outburst monitoring of the 14th luminous LMXB in a Galactic globular cluster, NGC 6440, the first cluster to show two luminous transient LMXBs (both of which show millisecond pulsations).  In a companion paper, \citet{Altamirano09} identify this LMXB as a new ultracompact accreting millisecond X-ray pulsar (AMXP), after discovering coherent 206 Hz pulsations \citep{Altamirano09b} and fitting the frequency drift to a 57.3 minute orbital period.

NGC 6440 is a globular cluster near the Galactic Center, at a distance of 8.5 kpc \citep{Ortolani94}, with $N_H=5.9\times10^{21}$ cm$^{-2}$ \citep{Harris96}.  A luminous X-ray source (MX 1746-20) was identified in this cluster in 1971 \citep{Markert75}, and the cluster was detected at a much lower flux level in 1980 \citep{Hertz83}, identified with quiescent emission from the luminous X-ray source.  An X-ray source in the cluster (SAX J1748.9-2021) was observed in outburst again in 1998 \citep{intZand99}, 2001 \citep{intZand01}, and 2005 \citep{Markwardt05}.  \citet{Pooley02b} observed the cluster with \Chandra\ in 2000, identifying 24 sources within 2 core radii of the cluster.  One (CX1) was positionally identified with a blue variable optical counterpart during the 1998 outburst \citep{Verbunt00}, and with the outbursting LMXB in 2001 \citep{intZand01}.  \citet{Altamirano07} identified intermittent 442 Hz pulsations in both the 2001 and 2005 outbursts (the latter also identified by \citealt{Gavriil07}). Thus, CX1 can be confidently held responsible for the 1998, 2001, and 2005 outbursts.  We here identify a second transient luminous LMXB in NGC 6440, and confidently identify five outbursts from this transient during 2009 using its source position.  Preliminary results were presented in \citet{Heinke09d,Heinke09f,Heinke09e}; these results supersede those.

\begin{figure}
\figurenum{1}
\includegraphics[angle=0,scale=.8]{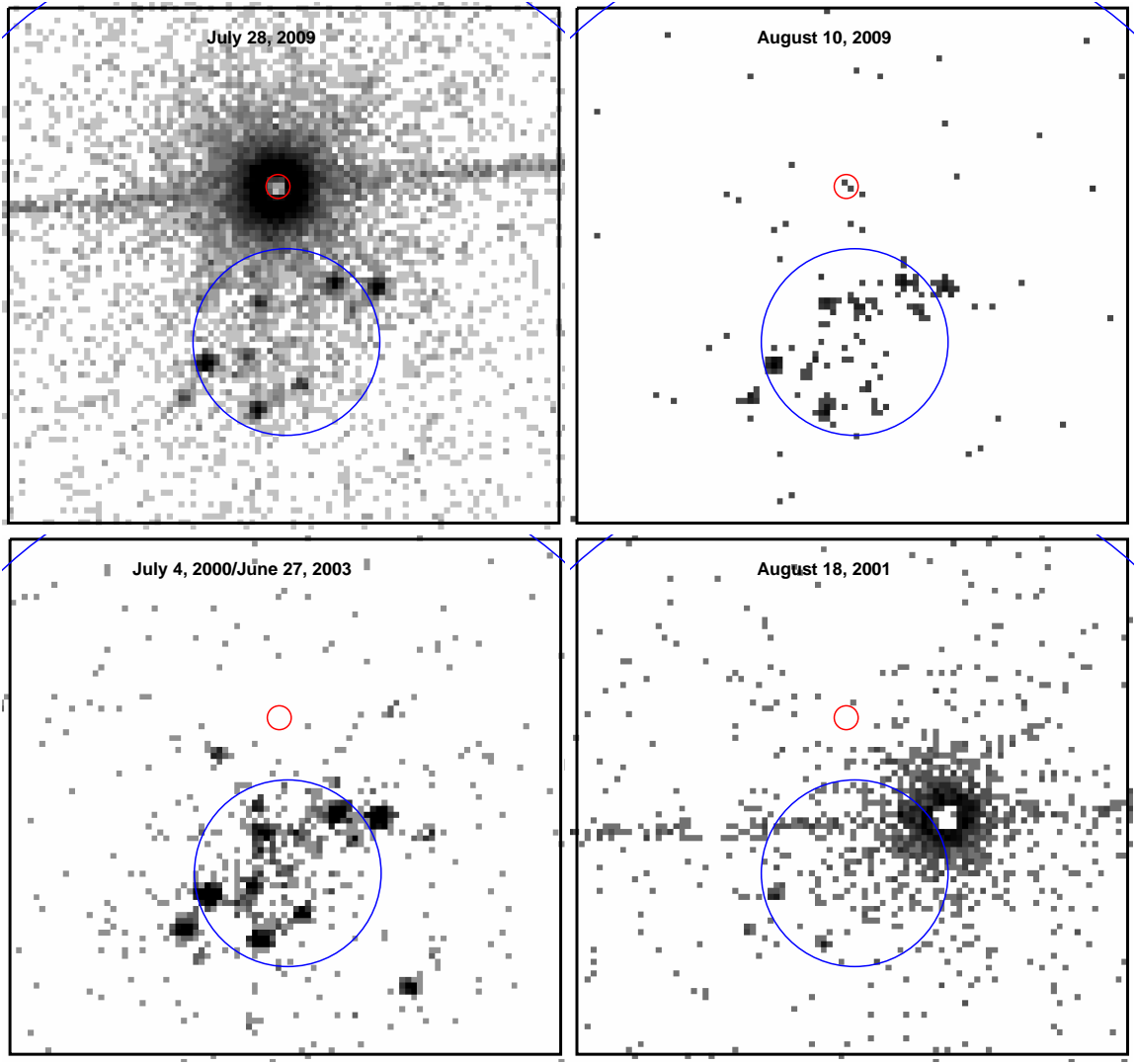}
\caption[6440-3images.ps]{ \label{fig:image}
\Chandra/ACIS-S images of NGC 6440 during the July 2009 outburst of NGC 6440 X-2 (upper left, 0.5-2.5 keV) and 13 days later (upper right, 0.3-7 keV).  Below, we show a merged image from quiescent epochs in 2000 and 2003 (lower left, 0.3-7 keV), and an image from the 2001 outburst (lower right, 0.3-7 keV). The positions of NGC 6440 X-2 (red circle, 1'') and the cluster core (0.13') and half-mass (0.58') radii (large circles) are indicated.  North is up, east to the left.
} 
\end{figure}
\clearpage
\section{X-ray Analysis}\label{s:X-ray}

\subsection{Chandra in Outburst} 
NGC 6440 was observed with the \Chandra\ ACIS-S detector for 49.1 ks, from July 28, 2009 15:16 (TT) to July 29, 2009 05:15, using a 1/2 subarray.  We searched for periods of elevated background, but found none.  We used the level 2 event files provided by the CXC, and CIAO 4.1 \footnote{http://cxc.harvard.edu/ciao} for our analysis.   Images were produced in the 0.3-7 keV and 0.5-2.5 keV bands, both of which are dominated by the scattered halo (a combination of dust grain scattering and the intrinsic point-spread function of the \Chandra\ mirrors) and readout streak from a bright transient LMXB, heavily affected by pileup\footnote{See the \Chandra\ Proposer's Observatory Guide, chapter 6.}.  Several faint point sources are clearly visible, which can be confidently identified with the cluster X-ray sources identified by \citet{Pooley02b}.  In Fig. \ref{fig:image}, we show \Chandra\ images of NGC 6440 during the 2009 outburst, shortly after the 2009 outburst, during quiescent observations in 2000 and 2003 \citep{Pooley02b,Cackett05}, and during the 2001 outburst \citep{intZand01}.  Clearly this is a new transient.

The CIAO detection algorithm {\tt wavdetect} was run on a 0.3-7 keV image of the cluster core to identify the positions of known cluster sources, which we shift ($\Delta$RA=+0.008s, $\Delta$Dec=+0.31\arcsec) to align with the (ICRS) astrometry of \citet{Pooley02b}.  We estimated (by eye) the center of the symmetric ``hole'' in the counts by matching circles to the doughnut-shaped locus of maximum countrate in the LMXB halo, in both wavebands.  Our result is (J2000) RA=17:48:52.76(2), Dec=-20:21:24.0(1) (1$\sigma$ values, after our astrometric correction), giving it the IAU name CXOGlb J174852.7-202124, and (for shorthand) NGC 6440 X-2.  Detailed analysis of the remaining cluster sources will be presented elsewhere.  

\begin{figure}[h]
\figurenum{2}
\includegraphics[angle=270,scale=.35]{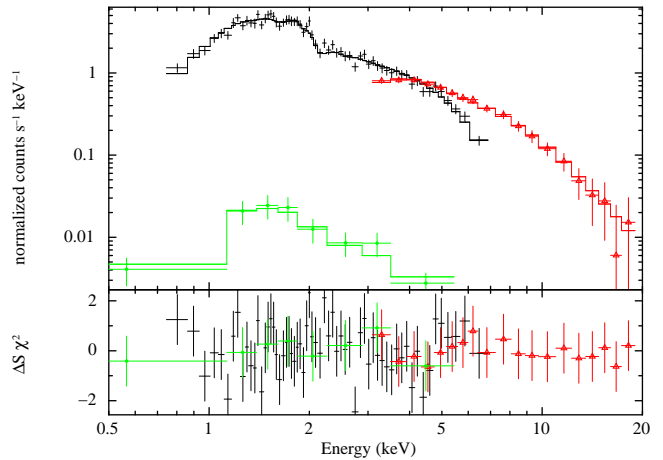}
\caption[X2spectra_bw.ps]{ \label{fig:spectrum}
 Spectra of NGC 6440 X-2 during the July/August outburst as observed by \Chandra/ACIS-S (black online), \RXTE/PCA (triangles; red online), and \Swift/XRT (filled circles; green online). Top: data and absorbed power-law models.  Bottom: residuals to fits.  See Table 1 for details of spectral fits.  
} 
\end{figure}

To measure the spectrum and luminosity of the transient, we extracted a spectrum from the readout streak, excluding a 20'' radius circle around the piled-up transient.  Background was  extracted from rectangular regions above and below the readout streak, computed response functions for the position of the transient, and corrected the exposure time of the spectrum.  We binned the spectrum to 60 counts per bin to improve its statistics, and excluded data over 8 keV and below 0.5 keV.  An absorbed power-law fits the data (Table 1, Fig. \ref{fig:spectrum}), with photon index 1.7$\pm0.1$.  We found no evidence for a 6.4 or 6.7 keV iron line, with a 90\% upper limit of 0.4 keV on its equivalent width.

The lightcurve from the readout streak events shows a clear decline during the Chandra observation, by a factor of $\sim$40\%.  
Power spectra from the readout streak events show a clear periodicity at 1000.0 seconds and its harmonics, identical to one of the spacecraft dither frequencies and thus a likely artifact (it is not seen in RXTE data).  No other periodicities (such as the 57 minute orbital period, Altamirano et al. 2009a) were identified, suggesting that the transient is not seen at high inclination. 

\subsection{RXTE} 
The {\it Rossi X-ray Timing Explorer} (\RXTE) has conducted regular scans of the Galactic bulge region since 2000 (generally covering NGC 6440 twice/week), in part to search for faint transient sources below the sensitivity of \RXTE's all-sky monitor \citep{Swank01}, with results promptly made available\footnote{ http://lheawww.gsfc.nasa.gov/users/craigm/galscan/main.html}.  The PCA scans are significantly more sensitive in the Galactic Bulge than the \RXTE\  All-Sky Monitor data.  (Although dwell-by-dwell All-Sky Monitor data suggest a few points at $L_X\sim10^{38}$ ergs/s from NGC 6440, suggesting X-ray bursts, their 6-sigma maximum significance is matched by similar points from black hole candidates in the Galactic Center.  Thus we cannot be assured of their reality.) Scans in late May 2009 showed evidence for an increased countrate from NGC 6440 (see 3rd red line in Fig. \ref{fig:lcurve}).  An outburst was confirmed by \Swift\ observations in early June (see below).  No further activity was observed until a scan showed a 5-sigma detection on July 28, 2009, near the beginning of the \Chandra\ observation.  A pointed \RXTE\ observation on July 30 showed a significant decline from the \Chandra\ and bulge scan fluxes (details below). 
Later bulge scans (on Aug. 1st \& 2nd) were consistent with a decay below the PCA's (background-limited, due to its non-imaging nature) sensitivity (Fig. \ref{fig:lcurve}).

 A third bulge scan peak on August 29, 2009 triggered a pointed observation Aug. 30, which discovered millisecond pulsations from NGC 6440 X-2 with a frequency of 205 Hz \citep{Altamirano09}.  Further \RXTE\ observations starting on Sept. 1st found fluxes returning to quiescence.  Upon identifying a fourth outburst with \Swift/XRT on Oct. 1st, 2009, a pointed \RXTE\ observation on Oct. 2nd detected NGC 6440 X-2 near the detection limit ($\sim$1.5 mCrab).  \citet{Altamirano09} report the detection of pulsations at a 3.4 sigma level, further confirming the identification of this transient with NGC 6440 X-2.  \RXTE\ observations on Oct. 3rd found it below the PCA's sensitivity limit.  A fifth outburst was caught by \RXTE/PCA on Oct. 28, returning to quiescence by Nov. 1st.
Below we describe spectral analysis of these data; all \RXTE/PCA timing analysis is described in \citet{Altamirano09}.

 We extracted spectra from the PCA \citep{Jahoda06} detectors (PCU2 was consistently on, sometimes joined by other PCUs), excluding times when elevation angle ELV$<10$\deg, source offset $>0.02$\deg, or the time since the last South Atlantic Anomaly passage was $<$ 30 minutes.  The background was modeled using the latest faint background model pca\_bkgd\_cmfaintl7\_eMv20051128.mdl, and the recorded times of South Atlantic Anomaly passage \footnote{http://heasarc.gsfc.nasa.gov/docs/xte/pca\_news.html}.  Response files were created with the PCARMF (v. 11.1) tool.  
For spectral fitting we exclude data below 3 keV and above where the source can no longer be detected above our modeled background; this cut ranged between 10 and 30 keV, depending on the observation.  HEXTE (cluster B, sensitive to the 20-200 keV energy range) did not detect the source even in the brightest (Aug. 30) pointed \RXTE\ observation. $N_H$ was fixed to the cluster value, due to \RXTE's relative insensitivity to $N_H$.  All our reported $L_X$ values are for 0.5-10 keV (for consistency with other instruments; this extrapolation may incur additional systematic errors), and assume a distance of 8.5 kpc.

The July 30 pointed observation could be fit with an absorbed power-law with photon index 2.2$\pm0.1$, and a luminosity of $5\pm1\times10^{35}$ \ergss, 3.5 times lower than the July 28 \Chandra\ observation (Table 1; but see below).  The August 30 observation showed a photon index of 1.84$\pm0.02$ and $L_X=2.9\times10^{36}$ \ergss.  The series of observations on Sept. 1 and 2 showed much lower fluxes, averaging $L_X=4\times10^{35}$ \ergss; the longest also shows clear evidence of an iron line at 6.7$\pm0.2$ keV.  However, simultaneous \Swift\ observations (see below) find much lower fluxes from the cluster of $6\times10^{34}$ \ergss, consistent with the quiescent cluster emission and inconsistent with the transient position.  Therefore we attribute the \RXTE/PCA flux observed on Sept. 1 to the Galactic Ridge emission at this location.  Subtracting this flux from the July 30 \RXTE/PCA measurement gives $L_X=1\pm1\times10^{35}$ \ergss, a very marginal detection, and makes its spectral parameters unreliable.   

We refine our spectral fitting of the August 30 observation, by extracting spectra from only the top layer of PCU2 and modeling the inferred Galactic Ridge emission using our fits to the Sept. 1st data.  Fitting with an absorbed powerlaw gives a good reduced chi-squared, but a residual near 6.5 keV suggests the addition of an iron line (this is in addition to the Galactic Ridge iron line seen in the Sept. data).  Freezing $N_H$ to the cluster value, we obtain $\Gamma=1.78\pm0.03$, $L_X$(0.5-10 keV)$=2.4\pm0.1\times10^{36}$ \ergss, with an iron line at $6.6^{+0.6}_{-0.4}$ keV, of equivalent width $54\pm39$ eV (90\% confidence).  An F-test suggests this is reasonable, providing 3.6\% probability of such a $\Delta \chi^2$ by chance. 
 Other AMXPs have displayed evidence of Fe K lines (e.g. SAX J1808.4-3658, \citealt{Cackett09,Papitto09}; HETE J1900.1-2455, \citealt{Cackett10}; Swift J1756.9-2508, \citealt{Patruno10}), often showing relativistic broadening which our observations are unable to resolve.

Two \RXTE/PCA observations were obtained during the early October outburst, both during the source's decline.  We identified pulsations in the first (Oct. 1st) observation, confirming that pulsations are common from this object \citep{Altamirano09}.  The second (Oct. 3rd) occurred after NGC 6440 X-2 had dipped below the PCA's sensitivity limit.  We model the Galactic Ridge contribution to the Oct. 1st observation, using the Oct. 3rd observation, finding $\Gamma=1.78\pm0.15$, $L_X$(0.5-10)$=8.4^{+0.8}_{-0.5}\times10^{35}$ \ergss (see Table 1).

Three \RXTE/PCA observations were able to detect the Oct./Nov. outburst.  The first (on Oct. 28) identified pulsations again \citep{Altamirano09}, at $L_X$(0.5-10 keV)$=8.6^{+0.9}_{-0.8}\times10^{35}$ ergs/s (see Table 1), after subtraction of Galactic Ridge emission.  \RXTE/PCA observations detected emission over background on Oct. 29 and (marginally) Oct. 30, and observations on Oct. 31 and after were consistent with Galactic Ridge emission.


\begin{figure}[h]
\figurenum{3}
\includegraphics[angle=0,scale=.45]{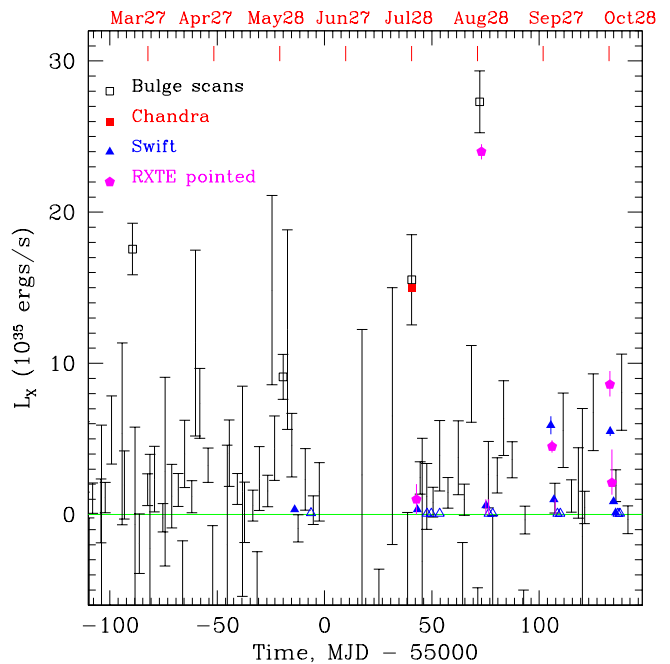}
\caption[trans_bw.eps]{ \label{fig:lcurve}
X-ray lightcurve of NGC 6440 over 6 months, with linear vertical scale.  
 \RXTE/PCA bulge scans are in black, with a box if they are $>4$ sigma above zero.   \RXTE/PCA pointed observations are magenta pentagons.  \Swift\ observations are blue triangles (open if upper limits).  The outburst \Chandra\ observation is a red filled box.  Red marks and dates at the top indicate the suggested 30.7 day recurrence epochs, which coincide with 5 clear episodes of X-ray activity.
} 
\end{figure}

The full \RXTE/PCA galactic bulge scan data reveals seven times from 2000 to 2009 when NGC 6440's count rates are 4 sigma above zero. The two brightest outbursts ($L_X>10^{37}$ \ergss) have been identified with the other transient in NGC 6440, SAX J1748.9-2021=CX1 \citep{Altamirano07,intZand01}.  The other five potential outbursts are much fainter and briefer (only a single bulge scan point each, so lasting less than a week), of which three were discussed above.  The other two are April 15, 2007 and March 20, 2009.  
A period of 30.7 days ($\pm$0.3 days, from the uncertainties in the peak of the late October outburst), with reference date MJD 55040.5, reasonably predicts the peaks of the four well-studied outbursts.  The model predicts missed outbursts on April 27 (weak activity is suggested by bulge scans on April 18-20), May 28 (a bulge scan detection occurred May 25, followed by a faint \Swift\ detection June 4), and June 27 (no bulge scans conducted $\pm$10 days around this date).  The bulge scan point in March is off the prediction by 7$\pm1$ days, indicating either that the outburst period has slightly decreased over time (as the April and May evidence also suggest), or that the March point is not a real outburst.  Sensitive monitoring since Nov. 2 (up to Feb. 11, 2010) has not been possible due to solar constraints and an outburst of SAX J1748.9-2021 \citep{Suzuki10, Patruno10b}.
We show the bulge scan data in the relevant date range (and detections and upper limits from \Swift\ and \Chandra) in Fig. \ref{fig:lcurve}; see also \citet{Altamirano09}.
We use an absorbed power-law of photon index 2 to convert \RXTE/PCA bulge scan fluxes to 0.5-10 keV unabsorbed fluxes with PIMMS\footnote{http://asc.harvard.edu/toolkit/pimms.jsp}, and assume a distance of 8.5 kpc to estimate luminosities.

\subsection{Swift} 
 Eighteen \Swift/XRT observations were performed, tracking five outbursts from NGC 6440 X-2 (see Table 1). 
We extracted all \Swift\ XRT spectra from 20-pixel radii (except for Oct. 1 and 28, see below), and background from a surrounding annulus.  We downloaded the response matrix  swxpc0to12\_20010101v009.rmf from the \Swift\ website\footnote{http://swift.gsfc.nasa.gov}, and created effective area files using the XRTMKARF tool.  Spectra with more than 50 counts were binned with 15 counts/bin for $\chi^2$ statistics, those with fewer used C-statistics (either binned with 5 counts/bin or unbinned), while we produced only luminosity limits for less than 10 detected counts. 

\begin{figure}[h]
\figurenum{4}
\includegraphics[angle=0,scale=.8]{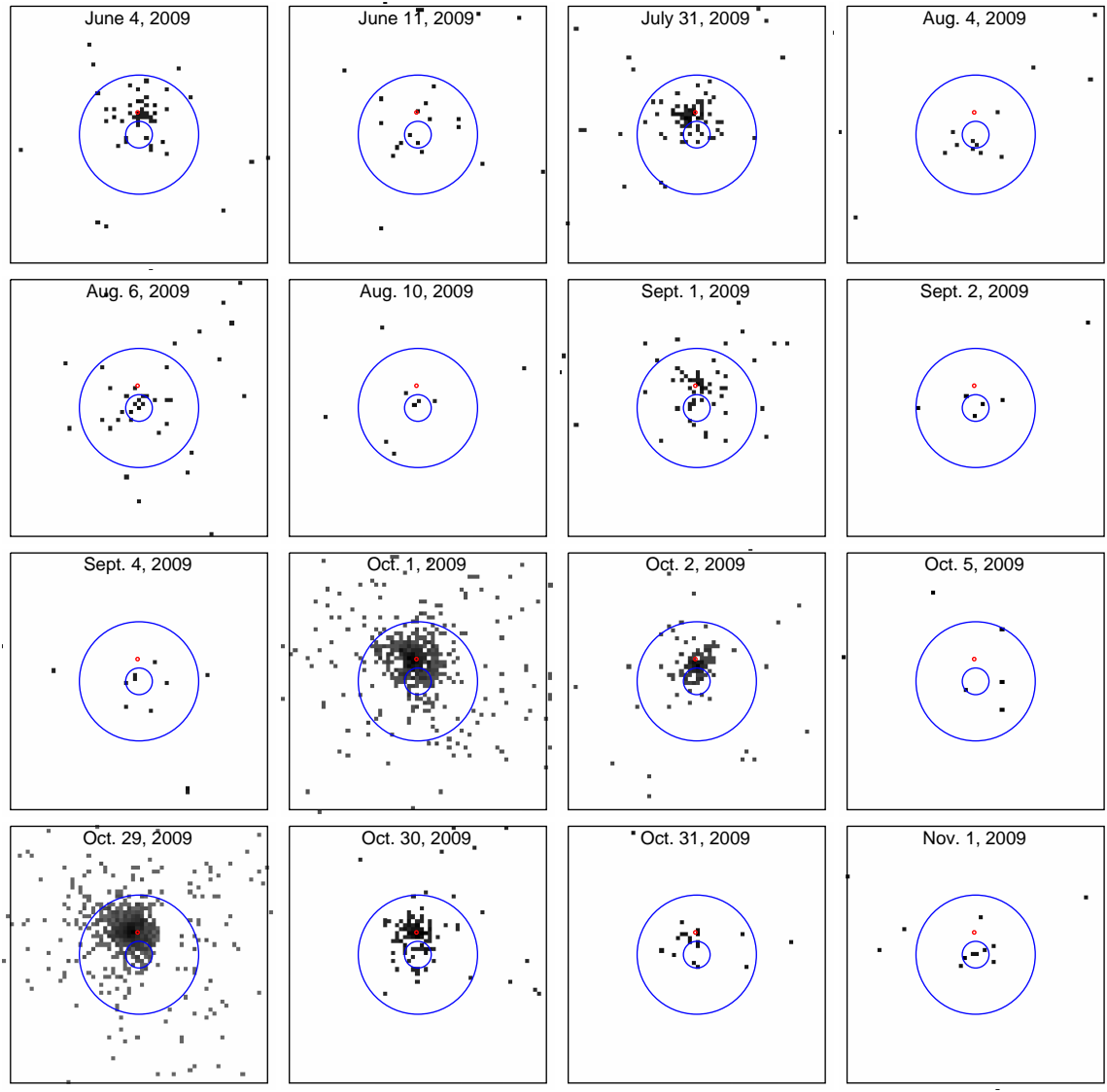} \\
\caption[Swift_all.ps]{ \label{fig:Swift6}
X-ray images (0.7-7 keV) of NGC 6440 during and after the five observed outbursts from the \Swift\ XRT, each 2.5' by 2.5'.  The position of NGC 6440 X-2 is indicated (small red circle), as are the core and half-mass radii of NGC 6440 (blue circles), as in Fig. \ref{fig:image}.  Exposure times vary (see Table 1).  Detections of NGC 6440 X-2 are seen on June 4, July 31, Sept. 1, Oct. 1, 2, 29, 30, and 31; the remaining emission is attributable to the other cluster sources (see Fig. \ref{fig:image}).
}
\end{figure}

A \Swift\ X-ray Telescope (XRT) observation on June 4, 2009 found enhanced X-ray emission from NGC 6440.
Using the FTOOL XRTCENTROID on the June 4 XRT source, we identified a position of RA=17:48:52.73, Dec=-20:21:24.1 with an error radius of 5\arcsec.  This position is consistent with NGC 6440 X-2 (see Fig. \ref{fig:image}, \ref{fig:Swift6}), but not with other known X-ray sources in NGC 6440, so we conclude it is the same source.  
A second \Swift\ observation on June 11 found much weaker emission, at position RA=17 48 52.64, Dec=-20 21 29.9, with error radius 7.3\arcsec, consistent with either NGC 6440 X-2 or with the cluster center.  Spectral fitting of the few detected photons with an absorbed power-law derives a photon index of 4.2$^{+1.2}_{-1.1}$ and $L_X=1.0^{+1.0}_{-0.5}\times10^{34}$ \ergss.  This is consistent with emission from the rest of the cluster, containing a mixture of soft quiescent LMXBs and sources (likely cataclysmic variables) with harder spectra \citep{Pooley02b,Heinke03d}.  

\begin{figure}[h]
\figurenum{5}
\begin{center} $
\begin{array}{cc}
\includegraphics[angle=0,scale=.4]{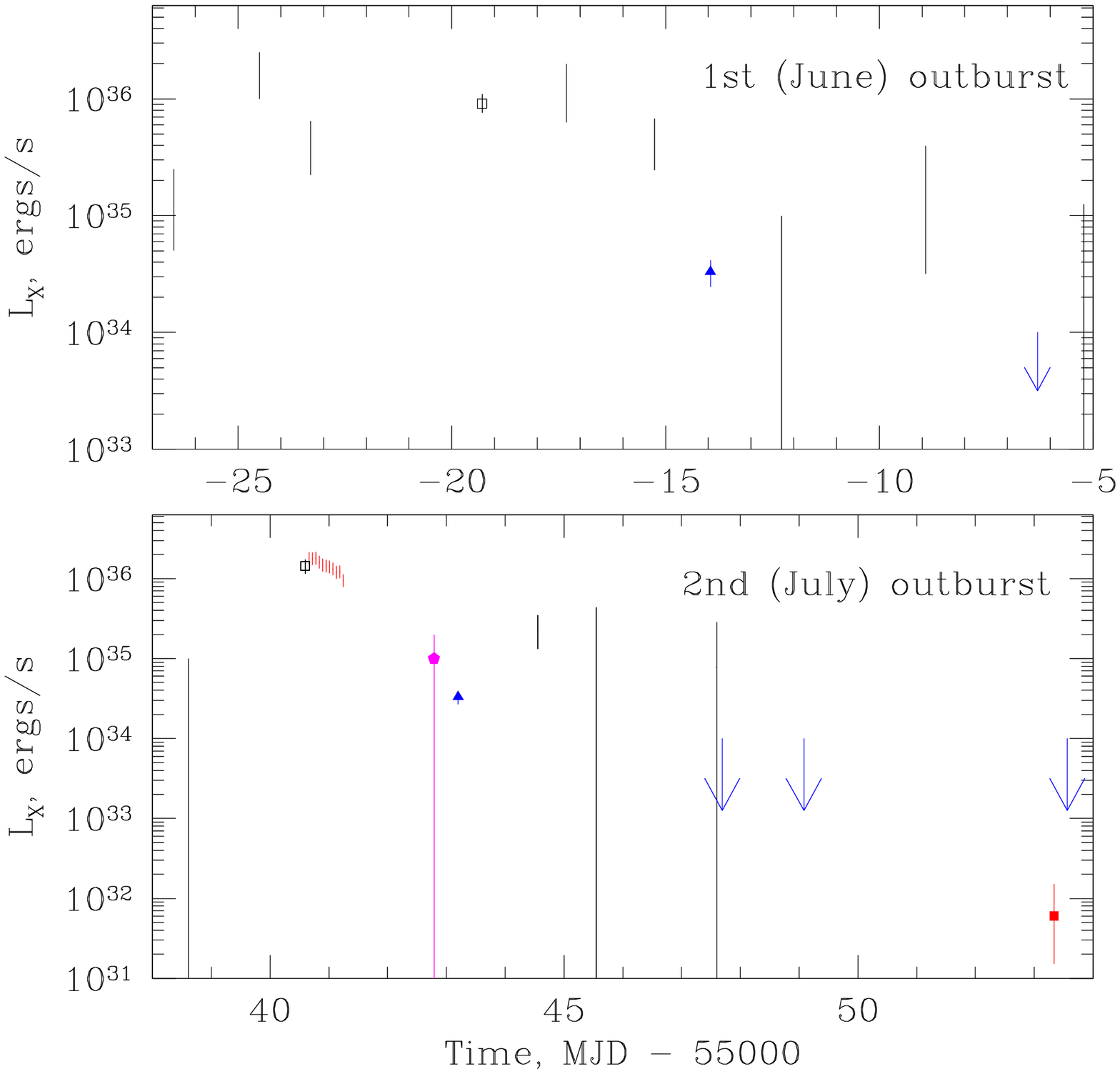} &
\includegraphics[angle=0,scale=.4]{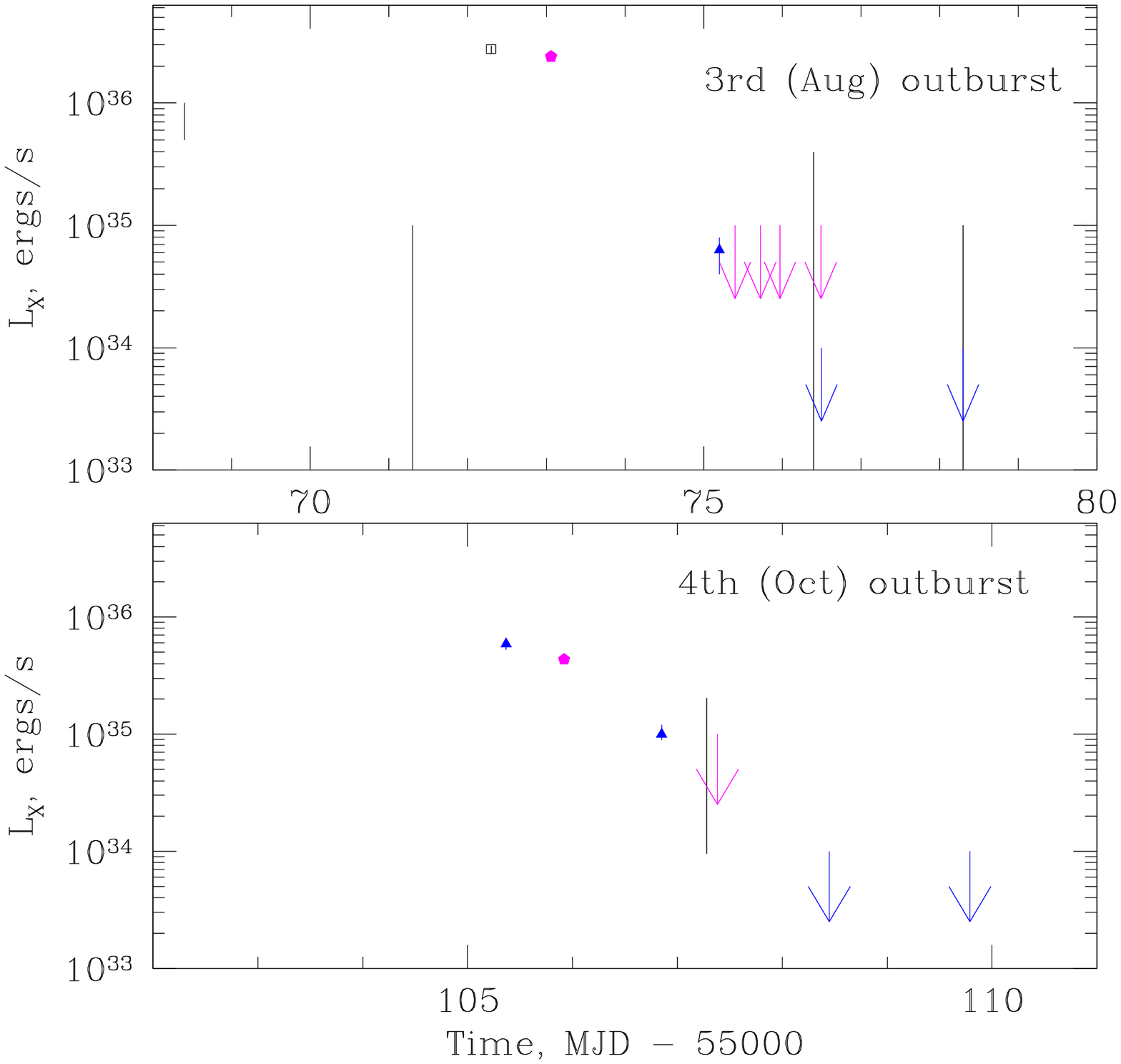}
\end{array}$
\end{center}
\includegraphics[angle=0,scale=.4]{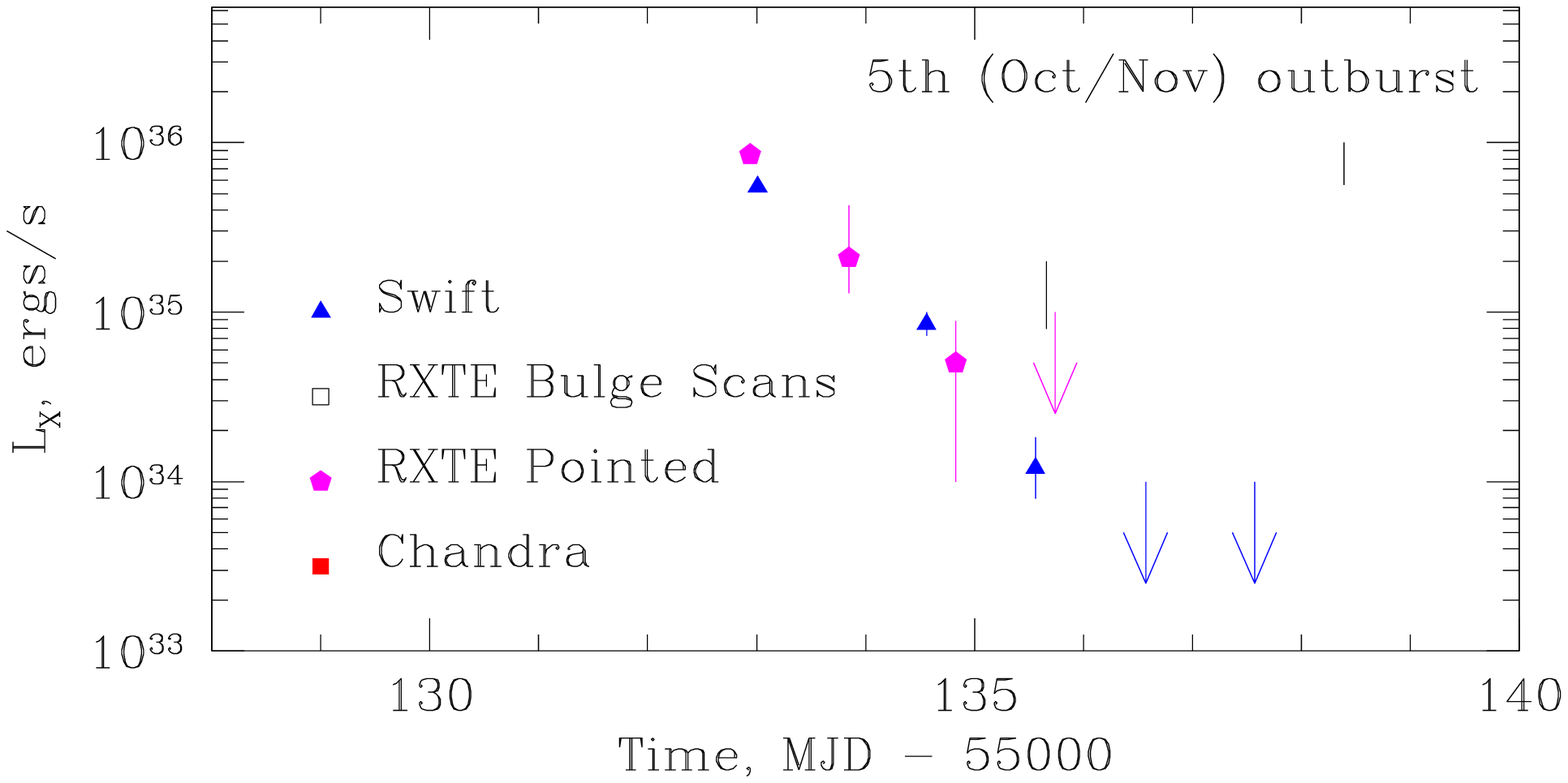}
\caption[inset_bw.eps]{ \label{fig:inset}
X-ray lightcurve measurements of NGC 6440 X-2 over its five known outbursts, with a logarithmic vertical scale.  \RXTE/PCA bulge scans are black errorbars, with a box point if they are $>4$ sigma above zero (some points appear more significant than they are due to the log scale).  \RXTE/PCA pointed observation detections (background-subtracted, see text) are magenta pentagons. \Swift\ observations are blue triangles, or blue upper limits when they are not clear detections of NGC 6440 X-2.  \Chandra\ observations are the red filled box and series of red crosses (showing the lightcurve from the June observation).
} 
\end{figure}

We obtained four \Swift\ observations soon after the initial \Chandra\ discovery (Table 1).  On July 31, \Swift/XRT found enhanced emission 
from NGC 6440 X-2's position. 
Observations on Aug. 4, 6, and 10 found fluxes and positions consistent with the cluster center, and soft spectra (Table 1, Figs. \ref{fig:Swift6}, \ref{fig:inset}).  The centroid of the emission in the deepest of these observations (Aug. 6, 1.9 ks) is RA=17:48:52.9s, Dec=-20:21:35.1, error radius 6.1'', which is consistent with the cluster center but not with NGC 6440 X-2. 

After the Aug. 29 bulge scan detection \Swift\ observed NGC 6440 on Sept. 1st, identifying emission from the location (RA=17:48:52.6, Dec=-20:21:24.9, error radius 4.8'') of NGC 6440 X-2 at $L_X=6\pm2\times10^{34}$ \ergss\ (Figs. \ref{fig:Swift6}, \ref{fig:inset}).  Further \Swift\ observations on Sept. 2 and 4 found faint emission, consistent in position and flux with the cluster.  

\Swift\ monitoring on Oct. 1st found NGC 6440 X-2 
back in outburst.  The countrate was high enough to produce pileup (in photon counting mode); we fit the radial profile with a King model and identified a 6'' core, so we extracted a spectrum from an annulus from 6'' to 45'' (20 pixels).  On Oct. 2nd, \Swift\ identified a declining flux from NGC 6440 X-2, returning to quiescence by Oct. 4th (Figs. \ref{fig:Swift6}, \ref{fig:inset}).  

A \Swift\ observation on Oct. 29 confirmed that NGC 6440 X-2 was back in outburst, and relatively bright; we dealt with pileup as described above.  Observations on Oct. 30 and 31st observed X-2's decay, which by Nov. 1 was below \Swift/XRT's detection limit  (Figs. \ref{fig:Swift6}, \ref{fig:inset}).

\clearpage

\clearpage

\subsection{Chandra in Quiescence}
We combined two \Chandra\ ACIS-S observations of NGC 6440 (total exposure 48 ks) when no transients were in outburst
 \citep[ObsIDs 947 and 3799; ][]{Pooley02b,Cackett05} to look for evidence of NGC 6440 X-2's X-ray emission in quiescence.  The observations were reprocessed and aligned, and filtered to produce a 0.3-7 keV band image (Fig. \ref{fig:image}).
No photons lie within a 1'' circle around NGC 6440 X-2.  Using 2.3 photons as our 90\% confidence upper limit \citep{Gehrels86}, we compute limits on the unabsorbed quiescent luminosity.  For a power-law of index 2.2, $L_X$(0.5-10 keV)$<7.4\times10^{30}$ \ergss, while for a hydrogen-atmosphere neutron star model  \citep{Heinke06a}, $L_X$(0.5-10 keV)$<1.6\times10^{31}$ \ergss and $L_{NS}$ (0.01-10 keV)$<6.0\times10^{31}$ \ergss (we use $L_{NS}$ as the total emission from the NS surface).  

We obtained a 5 ks ACIS-S follow-up \Chandra\ Director's Discretionary Time observation on Aug. 10, 2009, to see if NGC 6440 X-2 had returned to full quiescence, or was continuing to accrete at $L_X$(0.5-10 keV)$\sim5\times10^{32}$ \ergss, similarly to SAX J1808.4-3658 at the end of its 2008 outburst \citep{Campana08}.  Two photons were detected at the position of NGC 6440 X-2 (see Fig. \ref{fig:image}).  As the background is quite low ($<0.01$ photons expected), both photons are probably from NGC 6440 X-2.  We estimate (for an assumed spectrum similar to that in outburst)  $L_X=6^{+9}_{-4}\times10^{31}$ \ergss, with uncertainties from \citet{Gehrels86} at 90\% confidence.  This confirms that NGC 6440 X-2 returned to quiescence, although it  appears brighter than the limits from the 2000 and 2003 quiescent observations (above).  


\section{Optical/Infrared Observations}

\subsection{Archival NTT and HST}

The location of NGC 6440 X-2 has been previously observed by the European Southern Observatory's New Technology Telescope (NTT) and by the {\it Hubble Space Telescope} (\HST) \citep{Piotto02}.  The NTT imaging, in $R$ and $B$, was described by \citet{Verbunt00}, and the \HST\ imaging in $V$ and $B$ was described by \citet{Piotto02}.  We identified UCAC2 standards to place the NTT $R$ astrometric frame onto the ICRS frame, and then by identifying common stars in the WFPC2 and NTT frames, to place the \HST\ astrometric frame onto the ICRS frame, with an uncertainty of 0.2'' (1$\sigma$).  No star can be identified within 4$\sigma$ of the transient position on the NTT and WFPC2 $B$ frames, but one star in the WFPC2 $V$ frame is located 0.4'' from the transient position (Fig. \ref{fig:optical}).  Calibrating our photometry with that in \citet{Piotto02}, we find a magnitude of $V$=21.0$\pm0.2$ for this object.  Based on objects in the \HST\ image that are just barely detected in $B$, we estimate $B>22.0$.  These images were probably taken during quiescence, and NGC 6440 X-2's 57-minute orbital period indicates it will be very faint. As the ultracompact LMXB XTE J0929-314 in quiescence may have been detected with $M_V=13.2$ (D'Avanzo et al. 2009; thus it would have $V=31$ in NGC 6440), we think it unlikely that this star is the true counterpart.

\subsection{Outburst and Decay}

\citet{Li09} reported that unfiltered images were taken of NGC 6440 on July 30 and 31, 2009 with the 0.76-m Katzman Automatic Imaging Telescope (KAIT).  Their image subtraction against previous (2007, 2008) KAIT observations revealed no evidence of an optical transient, with limiting magnitudes of 19.5-20.0.

\begin{figure}[h]
\figurenum{7}
\includegraphics[angle=0,scale=.4]{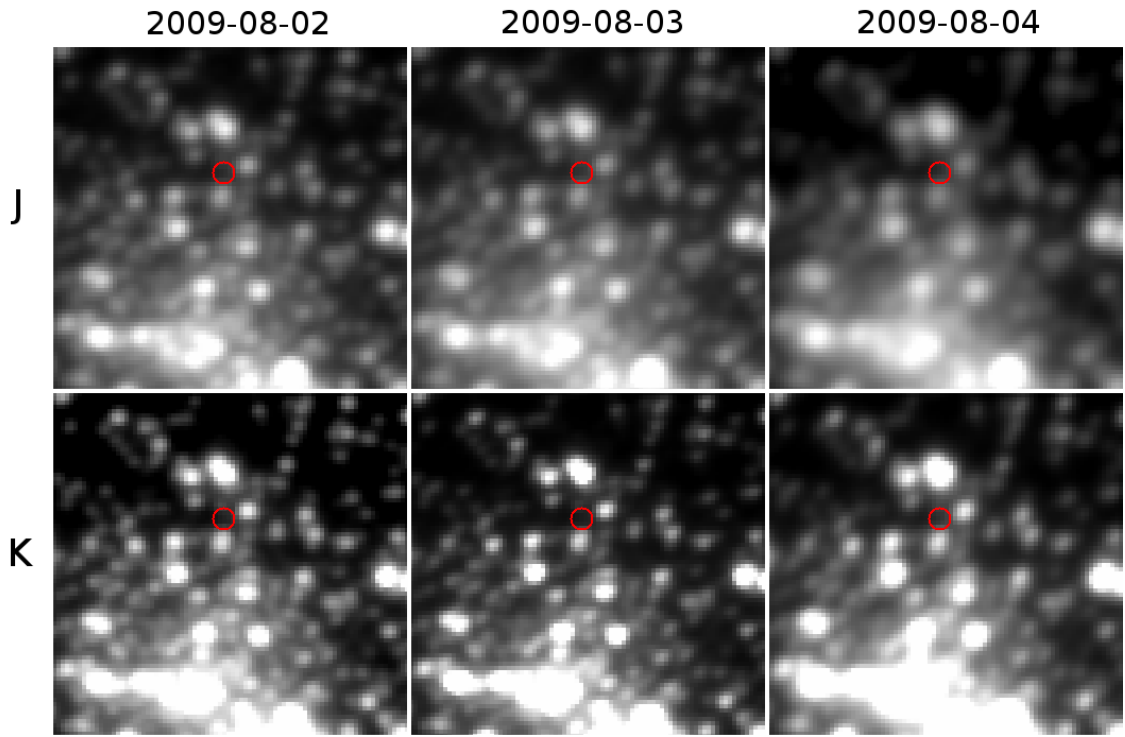}
\caption[optical]{ \label{fig:ISPI}
Infrared ISPI images (each 25''$\times$25'', N at top) with the CTIO 4-m on the dates specified.  Top, images in $J$; bottom, images in $K$.  Circles (0.8'' in radius) indicate the positions of NGC 6440 X-2.
}
\end{figure}

We obtained images with the CTIO 4-meter telescope using the ISPI infrared imager on Aug. 2-4, 2009 in the J and K bands (Fig. \ref{fig:ISPI}). The total exposure times were 12 min and 8 min per night, giving theoretical magnitude limits of 19.5 and 18 in J and K respectively (with a signal to noise ratio of 10). The airmass was around 1.1 and the seeing was 1'' the first night, increasing to 1.5'' the last night.
The images have been taken using a dithered pattern and an offset blank field has been observed to estimate the sky contribution to the emission. We used IRAF common packages and the PANIC package \citep{Martini04} to reduce the data, and then Sextractor/Scamp \citep{Bertin06} to calibrate the astrometry and photometry of the image, using the 2MASS catalog as a reference \citep{Skrutskie06}. Based on the detection of sources around NGC 6440 X-2, a source of magnitude K=17 and J=18.5 would have been detected.

\clearpage
\begin{figure}
\figurenum{6}
\includegraphics[angle=0,scale=.7]{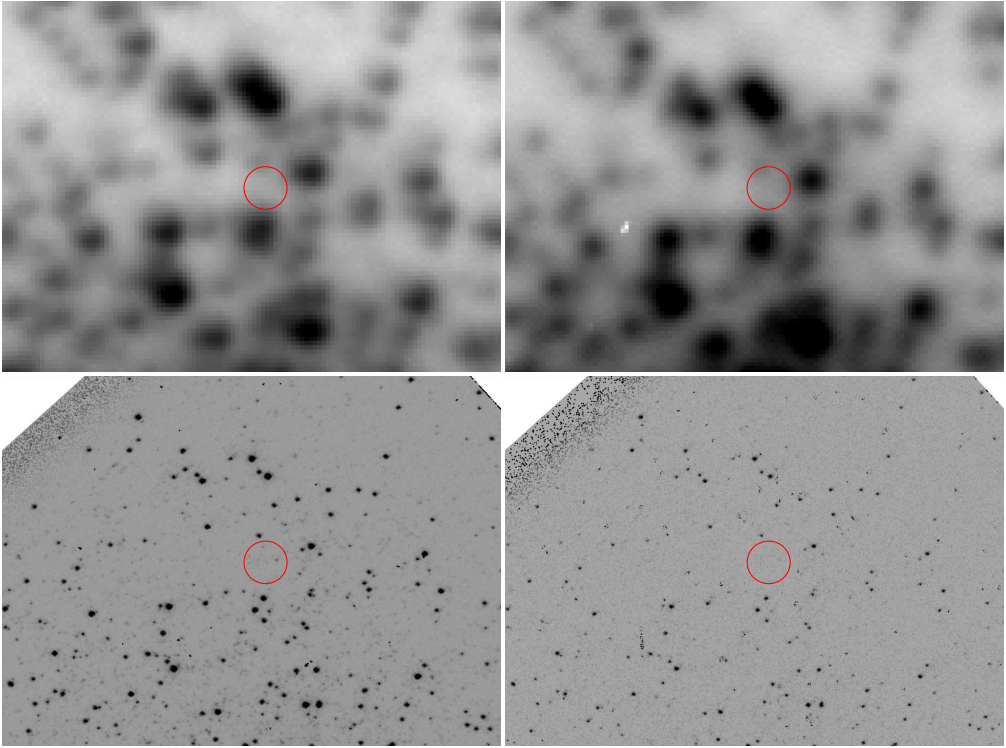}
\caption[optical]{ \label{fig:optical}
Optical images of NGC 6440, plotting an 0.8'' (4$\sigma$) error circle for NGC 6440 X-2.  Upper left, archival NTT $R$ frame; upper right, new Gemini $g'$ frame; lower left, \HST\ WFPC2 $V$ frame; lower right, \HST\ WFPC2 $B$ frame.  All 19'' $\times$ 14'', with N at top.
}
\end{figure}

\clearpage

We obtained three 500 s frames in each of $g'$ (note that the $g'$ bandpass is between $V$ and $B$) and $r'$ using Gemini-South GMOS-S on Sept. 1, 2009, during the outburst decay (program GS-2009B-DD-2).  Unfortunately the $r'$ frames were saturated at the location of the transient, but the $g'$ frames provided our best optical outburst limit (Fig. \ref{fig:optical}).  No star was seen within 0.8\arcsec ($4\sigma$) of the transient position. 
An approximate calibration of the $g'$ images was obtained using a star by star comparison with the $B,~V$ photometry of \citet{Martins80}.  We determined the detection limit for stars within an annulus 
about the cluster center containing the location of the transient.  
Based on the faintest 5\% of the stars in the annulus, 
the estimated detection limit is $g' \sim 22.0$.  
Thus we adopt $g'>22$ as an approximate limit for NGC 6440 X-2, at a time when \Swift\ found its X-ray luminosity to be $6\times10^{34}$ \ergss.

\section{Discussion}\label{s:discuss}

The discovery of millisecond pulsations \citep{Altamirano09} clearly identifies NGC 6440 X-2 as a neutron star LMXB system, and provides the orbital period of 57.3 minutes.  
The ultracompact nature of NGC 6440 X-2 confirms the trend \citep{Deutsch00} for globular clusters to have a higher fraction of ultracompact LMXBs than field systems, indicating a different formation mechanism \citep{Ivanova05}.  However, it goes against the apparent trend \citep{Zurek09} for ultracompacts in globular clusters to have shorter periods than ultracompacts in the field.  
The lack of optical detections is not surprising, when the ultracompact orbit is considered.  Using the relation between absolute magnitude, orbital period, and X-ray luminosity derived by \citet{vanParadijs94}, we predict $M_V=3.8$ and thus $V=21.8$ at peak.  The scatter in this relation is about 1.5 magnitudes, and our optical observations all occurred during the outburst decays, so our limits on NGC 6440 X-2 are consistent with this relation.

NGC 6440 X-2's outbursts are unusual among globular cluster LMXBs, both for the faintness of the outbursts (peak 0.5-10 keV $L_X=1.5\times10^{36}$ \ergss\ for the July outburst, $2.8\times10^{36}$ \ergss\ for the August outburst), but more importantly for their brevity.  The July X-ray lightcurve indicates that the time spent above $10^{35}$ \ergss\ was no more than 4 days, and perhaps only 2.5 days (Fig. \ref{fig:inset}), one of the shortest transient LMXB outbursts so far recorded \citep[cf.][]{Natalucci00,intZand04,Wijnands07}.  
The August outburst lightcurve indicates $<3.5$ days spent above $10^{35}$ \ergss, and the lightcurves from the other outbursts, while less constraining, are consistent with this timescale (Fig. \ref{fig:inset}).  We note that another ultracompact AMXP, XTE J1751-305, has shown  similarly short and faint outbursts \citep{Markwardt07,Linares07,Markwardt09}.  

Why does NGC 6440 X-2 show such faint and brief outbursts?  The obvious drivers are accretion disk instabilities or magnetospheric instabilities.  However, magnetospheric instabilities occur on much faster timescales (e.g. the Rapid Burster, Lewin 1993), and would require that the disk remain viscous (and thus ionized) between outbursts, which seems unlikely.  Standard accretion disk instability models (e.g. Lasota 2001) predict quiescent periods 10 times longer, and brighter and longer outbursts, for systems with 57-minute orbital periods.  Naively, we expect longer intervals between outbursts for a 57-minute system than for the other known ultracompact systems, if NGC 6440 X-2 is indeed an evolutionary descendant of systems like them, as it will have a larger disk and lower mass transfer rate (e.g. Deloye \& Bildsten 2003).   However, we are not aware of detailed modeling of outbursts of hydrogen-poor accretion disks with extreme mass ratios (and thus likely superhumps, Whitehurst 1988), suggesting an avenue for further study. 

  The X-ray record suggests that NGC 6440 X-2's activity has significantly increased in the past year.  It would be difficult to attribute this change to the disk (as it includes numerous outburst cycles), and we suggest that it represents a signal of mass-transfer variations from the donor.  Cyclical variations in the orbital period are well-established in longer-period cataclysmic variables \citep{Borges08}, LMXBs \citep{Wolff09}, and black widow pulsars \citep{Arzoumanian94} and seem likely to be due to magnetic activity in the companion star.  Such activity has also been suggested to explain SAX J1808.4-3658's large current rate of orbital period increase \citep{Hartman08}, and decades-long variations in mass transfer rates in LMXBs \citep{Durant09}.  If such a mechanism is active here, it requires a partly nondegenerate, convective companion.  Alternatively, the increased activity could indicate a change in the orbital parameters, induced by a distant companion (making this a triple system; testable with monitoring of future outbursts by RXTE) or a recent close interaction with another star (its position outside the core suggests this is less likely).

It is difficult to estimate NGC 6440 X-2's mass transfer rate, due to its extreme faintness and the limitations of existing surveys.  We identify two limiting cases, one based on its recent outburst history, and one using the full bulge scan light curve.  
For the first case, we estimate the outbursts as lasting 3 days at an average $L_X\sim1.5\times10^{36}$ \ergss, and occurring every 31 days.
For canonical neutron star mass and radius estimates, this gives a time-averaged mass accretion rate of $3\times10^{-11}$ \Msun/year.  Although this represents the mass transfer rate over the past few months, it is clear that NGC 6440 X-2 has not shown such outbursts regularly over the entire bulge scan epoch, where only 5 outbursts have been identified.  Assuming (generously) that 2/3 of all outbursts have been missed (the October and November outbursts were missed by bulge scans, and it seems likely that outbursts in June/July and April/May were missed; Fig. \ref{fig:lcurve}), and that the average outburst is like those seen so far, we estimate NGC 6440 X-2's mass transfer rate over the entire bulge scan epoch (10 years) as  $1.3\times10^{-12}$ \Msun/year.  This latter rate is consistent with an ultracompact binary of orbital period 57 minutes experiencing conservative mass transfer driven by general relativistic angular momentum loss \citep{Deloye03}, though a higher rate is not inconsistent with a relatively high-entropy (partly nondegenerate) donor.

\begin{figure}[h]
\figurenum{8}
\includegraphics[angle=0,scale=.45]{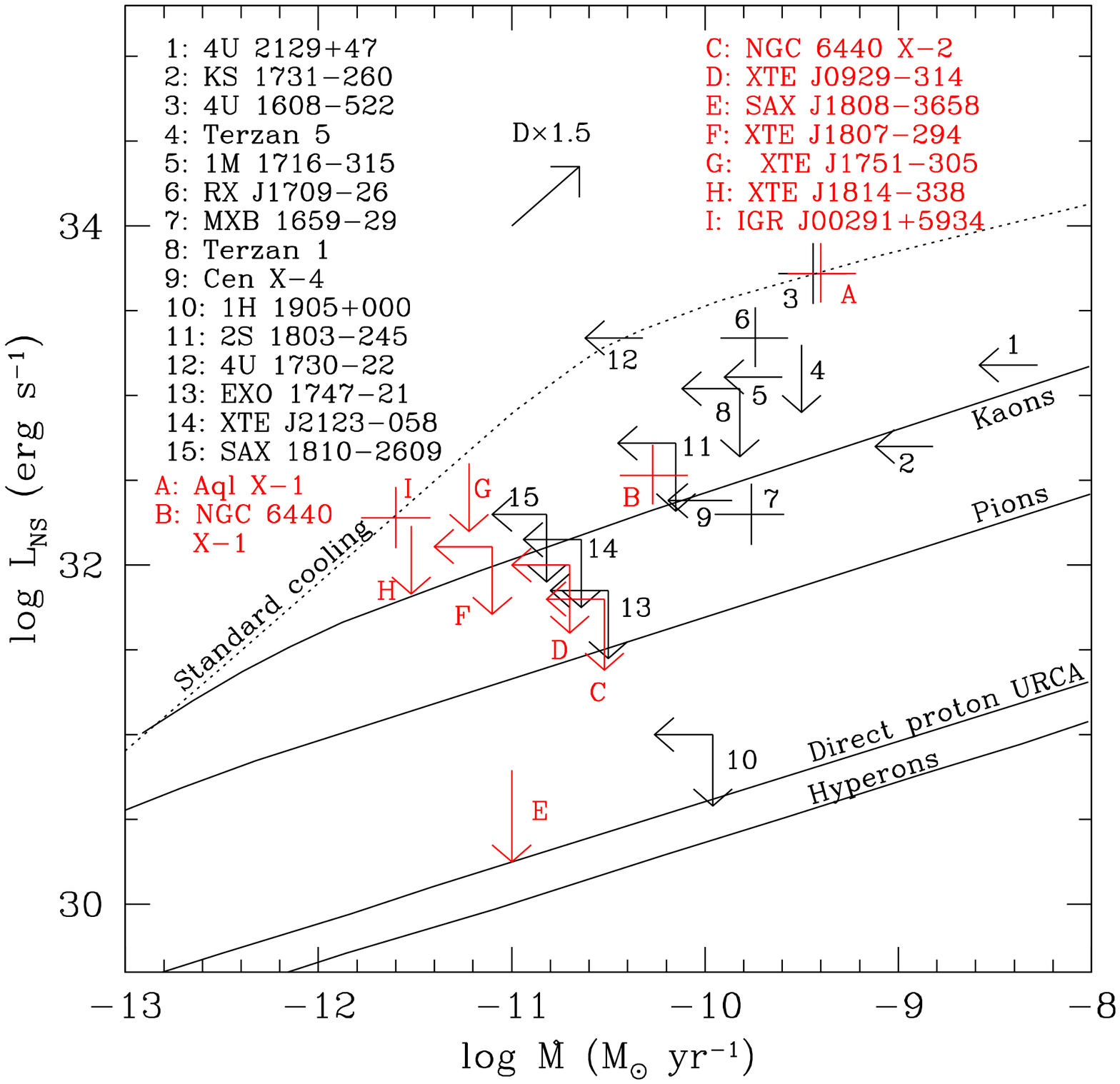}
\caption[optical]{ \label{fig:yak_rev}
Measurements of, or limits on, the quiescent thermal luminosity of various NS transients, compared to estimates of, or upper limits on, their time-averaged mass accretion rates.  Data from compilations of \citet{Heinke07a,Heinke09a}, with NGC 6440 X-2 added.  Predictions of standard cooling and several enhanced cooling mechanisms are plotted, following \citet{Yakovlev04}.  Accreting millisecond pulsars are indicated separately (in red), while the effect of increasing the distance by a factor of 1.5 for any system is indicated with an arrow labeled ``D$\times$1.5''.
}
\end{figure}

 The tight upper limit on NGC 6440 X-2's quiescent emission is the third lowest for any neutron star LMXB, after the transients SAX J1808.4-3658 and 1H 1905+000 \citep[][; Fig. \ref{fig:yak_rev}]{Heinke09a,Jonker07}.  Deep \Chandra\ observations might substantially improve these limits (e.g. 100 ks could reduce the quiescent flux limit by a factor of 3).
Long-term study of outbursts from this system will allow a better measure of the average mass accretion rate.  It will be of great interest to see if the outbursts continue to occur every $\sim$31 days, turn off, or change their outburst frequency, as this system's behavior is extremely unusual.

This is the first globular cluster to show two transiently outbursting X-ray sources.  Many candidate quiescent LMXBs have been identified in globular clusters through their soft spectra, including eight in NGC 6440 \citep{Grindlay01a,Rutledge02a,Pooley02b, Heinke03d}, although few have been observed to undergo outbursts.  Some of these quiescent LMXBs may be producing short, faint transient outbursts like NGC 6440 X-2's, which are at or near the noise level for existing surveys such as the \RXTE/PCA bulge scans and All-Sky Monitor.  Even fainter X-ray transients have been studied in the Galactic Center with dedicated observations \citep{Muno05,Wijnands06}.   \Swift\ could efficiently survey one or a few of the globular clusters richest in quiescent LMXBs for such small-scale outbursts.  

\acknowledgements

We are grateful to N. Gehrels and the \Swift\ team, H. Tananbaum and the \Chandra\ team, M. Pretorius at ESO, M. Buxton at SMARTS, the \RXTE\ team, N. Levenson, J.  Radomski, R. Carrasco, and the Gemini-South Science Team, for rapidly scheduling observations of NGC 6440.  We thank 
D. Pooley, S. Ransom, N. Degenaar, \& A. Kong for discussions, and the referee for a useful, clear and rapid report.
 This research has made use of data obtained through the High Energy Astrophysics Science Archive Research Center Online Service, provided by the NASA/Goddard Space Flight Center.  We acknowledge the use of public data from the \Swift, \RXTE, \Chandra, {\it HST}, and ESO data archives.  

{\it Facilities:} \facility{RXTE (PCA)}, \facility{CXO (ACIS)}, \facility{Swift (XRT)}, \facility{Gemini:South (GMOS)}, \facility{HST (WFPC2)}, \facility{NTT ()}

\bibliography{src_ref_list}
\bibliographystyle{apj}


\begin{deluxetable}{cccccccc}
\setlength{\tabcolsep}{1pt}
\tablewidth{7.0truein}
\tablecaption{\textbf{Spectral Fits to NGC 6440 X-2}}
\tablehead{
{\textbf{Start time}} & {\textbf{Instrument}} & {Exposure} & {$N_H$} & {$\Gamma$} &  {$\chi^2_{\nu}$/dof} & {$L_X$, 0.5-10 keV} & {ObsID} \\
 (2009) & & ks & $\times10^{21}$ cm$^{-2}$ & & & \ergss &   \\}
\startdata
June 4, 01:04 & \Swift/XRT & 1.1 & 0.59$^a$ & 2.4$\pm0.5$ & 49$^b$ & $3.3\pm0.8\times10^{34}$ & 00031421001 \\
June 11, 17:32 & \Swift/XRT & 1.9 & 0.59$^a$ & 4.2$^{+1.2}_{-1.1}$ & 59$^b$ & $1.0^{+1.0}_{-0.5}\times10^{34}$ & 00031421002 \\
\hline
July 28, 20:17 & \Chandra/ACIS-S & 49.1 & 0.69$\pm0.06$ & 1.79$\pm0.10$ & 0.98/59 & $1.55\pm0.06\times10^{36}$ & 10060 \\
July 30, 18:47 & \RXTE/PCA &    1.9  & 0.59$^a$ & 2.1$^{+0.8}_{-0.7}$ & 0.57/27 & $9.5^{+13}_{-4.5}\times10^{34}$ & 94044-04-01-00 \\
July 31, 04:55 & \Swift/XRT & 1.8 & 0.59$^a$ & 2.3$\pm0.4$ & 92$^b$ & $3.1\pm0.6\times10^{34}$ & 00031421003 \\
Aug. 4, 16:26  & \Swift/XRT & 1.0 & 0.59$^a$ & 2.9$^{+1.2}_{-1.1}$ & 37$^b$ & $7^{+8}_{-3}\times10^{33}$ & 00031421004 \\
Aug. 6, 01:50  & \Swift/XRT & 1.9 & 0.59$^a$ & 3.4$^{+1.2}_{-1.0}$ & 99$^b$ & $1.1^{+0.9}_{-0.5}\times10^{34}$ & 00031421005 \\
Aug. 10, 13:28 & \Swift/XRT & 0.5 & 0.59$^a$ & $^c$ & - & $7^{+8}_{-5}\times10^{33}$ & 00031421007 \\
Aug. 10, 08:10 & \Chandra/ACIS-S & 4.9 & 0.59$^a$ & $^c$ & - & $6^{+9}_{-5}\times10^{31}$ & 11802 \\ 
\hline
Aug. 30, 1:31$^d$ & \RXTE/PCA & 3.2 & 0.59$^a$ & 1.79$\pm0.02$ & 0.84/54 & $2.55\pm0.05\times10^{36}$ & 94044-04-02-00 \\ 
Sept. 1,  3:43$^d$ & \RXTE/PCA & 14.1 & 0.59$^a$ & $1.5^{+0.8}_{-0.8}$ & 0.8/33 & $<4\times10^{34}$  & 94044-04-02-01 \\  
Sept. 1, 4:33 &   \Swift/XRT & 1.0 & 0.59$^a$ & $3.8^{+0.7}_{-0.6}$ & 90$^b$ & $6\pm2\times10^{34}$  & 00031421009 \\
Sept. 2, 13:01 & \Swift/XRT  & 0.4  & 0.59$^a$  &  $^c$ & -  & $1.2^{+1.1}_{-6}\times10^{34}$ &  00031421010 \\
Sept. 4, 06:49 & \Swift/XRT  &   0.4  & 0.59$^a$ &  $^c$  &  - & $1.5^{+1.3}_{-0.7}\times10^{34}$ &  00031421012 \\
\hline 
Oct. 1, 08:47  & \Swift/XRT & 0.7 & 0.59$^a$ & $1.79\pm0.19$ & 1.0/16 & $5.9\pm0.6\times10^{35}$ & 00031421014 \\
Oct. 1, 21:58 & \RXTE/PCA & 1.5 & 0.59$^a$ & 1.8$^{+0.1}_{-0.2}$ & 0.49/29 & $4.5^{+0.4}_{-0.4}\times10^{35}$ & 94044-04-03-00 \\ 
Oct. 2, 20:24 & \Swift/XRT & 1.0 & 0.59$^a$ & 1.9$\pm0.3$ & 0.35/6 & $1.0^{+0.2}_{-0.1}\times10^{35}$ & 00031421015 \\
Oct. 3, 08:12 & \RXTE/PCA & 3.2 & 0.59$^a$ & - & 1.1/28 & $<3\times10^{34}$ & 94044-04-04-00 \\ 
Oct. 4, 10:46 & \Swift/XRT & 0.6 & 0.59$^a$ & $^c$ & - & $6^{+6}_{-4}\times10^{33}$ &  00031421017 \\
Oct. 5, 18:59 & \Swift/XRT & 0.6 & 0.59$^a$ & $^c$ & - & $7^{+7}_{-4}\times10^{33}$ & 00031421016 \\
\hline
Oct. 28, 22:29 & \RXTE/PCA & 0.9 & 0.59$^a$ & 1.8$^{+0.1}_{-0.1}$ & 0.34/39 & $8.6^{+0.9}_{-0.8}\times10^{35}$ & 94315-01-04-01 \\
Oct. 29, 00:19 & \Swift/XRT & 1.2 & 0.59$^a$ & 1.7$^{+0.1}_{-0.1}$ & 1.03/40 & $5.5^{+0.3}_{-0.3}\times10^{35}$ & 00031421018 \\
Oct. 29, 20:25 & \RXTE/PCA & 0.9 & 0.59$^a$ & 2.2$^{+0.6}_{-0.5}$ & 0.58/27 & $2.1^{+2.2}_{-0.8}\times10^{35}$ & 94315-01-04-02 \\	
Oct. 30, 13:28 & \Swift/XRT & 0.9 & 0.59$^a$ & 2.3$^{+0.3}_{-0.4}$ & 1.15/8 & $8.6^{+1.4}_{-1.4}\times10^{34}$  & 00031421019 \\
Oct. 30, 19:58 & \RXTE/PCA & 0.8 & 0.59$^a$ & $3.0^{+4.4}_{-2.3}$ & 0.52/15 & $5\pm4\times10^{34}$ & 94315-01-03-01 \\
Oct. 31, 13:33 & \Swift/XRT & 0.9 & - & $^c$ & - & $1.2^{+0.6}_{-0.4}\times10^{34}$ & 00031421020 \\
Oct. 31, 17:48 & \RXTE/PCA & 1.1 & 0.59$^a$ & (2) & 0.58/23 & $<3.1\times10^{34}$ & 94315-01-03-02 \\
Nov. 1, 13:39 & \Swift/XRT & 0.9 & - & $^c$ & - & $6^{+4}_{-3}\times10^{33}$ & 00031421021 \\
Nov. 2, 13:45 & \Swift/XRT & 0.9 & - & $^c$ & - & $4^{+4}_{-2}\times10^{33}$ & 00031421022 \\
\enddata
\tablecomments{Absorbed power-law spectral fits, using distance of 8.5 kpc.   
Errors are 90\% confidence for a single parameter.  
\Swift\ results apply to flux from entire cluster (see text for details), while \RXTE\ results 
subtract the quiescent \RXTE/PCA fitted spectrum (inferred $L_X=4\times10^{35}$ ergs/s) produced by the Galactic Ridge.  After an \RXTE/PCA observation finds only an upper limit during the decline of an outbursts, we omit the later \RXTE/PCA observations from this table (omitted ObsIDs are 94044-04-02-13, 94044-04-02-02, 94044-04-02-03, 94044-05-01-00, 94315-01-03-03, 94315-01-03-04, 94315-01-03-09,  94315-01-03-05, 94315-01-03-06, 94315-01-03-00, \& 94315-01-03-07.).
$^a$:$N_H$ is held fixed at cluster value. 
$^b$:Statistics low, C-statistic used and goodness reported.
$^c$:Too few counts ($<10$) detected for any spectral fitting; see text.
$^d$: Including an Fe line at 6.6 keV, assumed width 0.1 keV; see text for details.
}
\end{deluxetable}


\end{document}